\documentclass[final]{agujournal2019}
\usepackage{url} %this package should fix any errors with URLs in refs.
\usepackage{lineno}

\usepackage[finalnew]{trackchanges}
\usepackage{soul}
\usepackage[version=4]{mhchem}
\usepackage{amsmath, amssymb}
\usepackage{tabularx}
\usepackage{booktabs}
\draftfalse

\let\oldequation\equation
\let\oldendequation\endequation

\renewenvironment{equation}
  {\linenomathNonumbers\oldequation}
  {\oldendequation\endlinenomath}

\journalname{JGR: Planets}

\begin{document}
%TC:ignore
%% ------------------------------------------------------------------------ %%
%  Title
%

\title{Growth and evolution of secondary volcanic atmospheres: I. Identifying the geological character of hot rocky planets}

%% ------------------------------------------------------------------------ %%
%
%  AUTHORS AND AFFILIATIONS
%
%% ------------------------------------------------------------------------ %%

% Authors are individuals who have significantly contributed to the
% research and preparation of the article. Group authors are allowed, if
% each author in the group is separately identified in an appendix.)

% List authors by first name or initial followed by last name and
% separated by commas. Use \affil{} to number affiliations, and
% \thanks{} for author notes.
% Additional author notes should be indicated with \thanks{} (for
% example, for current addresses).

\authors{Philippa Liggins\affil{1}, Sean Jordan\affil{2}, Paul B. Rimmer\affil{1}\affil{3}\affil{4}, Oliver Shorttle\affil{1}\affil{2}}

\affiliation{1}{Department of Earth Sciences, University of Cambridge, United Kingdom} % Downing St, Cambridge CB2 3EQ,
\affiliation{2}{Institute of Astronomy, University of Cambridge, United Kingdom} %  Madingley Rd, Cambridge CB3 0HA,
\affiliation{3}{Cavendish Laboratory, University of Cambridge, United Kingdom} % , JJ Thomson Ave, Cambridge CB3 0HE
\affiliation{4}{MRC Laboratory of Molecular Biology, Francis Crick Ave, Cambridge, United Kingdom} %  CB2 0QH

%% Corresponding Author:

\correspondingauthor{Philippa Liggins}{pkl28@cam.ac.uk}

%% Keypoints, final entry on title page.

\begin{keypoints}
\item Varying the mantle $f\mathrm{O}_2$ of hot rocky planets can produce three distinct atmospheric classes
\item The planet's mantle $f\mathrm{O}_2$, H/C ratio and age control whether the main atmospheric species is \ce{H2O}, \ce{H2} or \ce{CO2}
\item The chemical speciation of volcanic atmospheres is robust to a wide range of bulk silicate planet H/C mass ratios
\end{keypoints}
%TC:endignore
%% Abstract

\begin{abstract}
The geology of Earth and super-Earth sized planets will, in many cases, only be observable via their atmospheres. Here, we investigate secondary volcanic atmospheres as a key base case of how atmospheres may reflect planetary geochemistry. We couple volcanic outgassing with atmospheric chemistry models to simulate the growth of C-O-H-S-N atmospheres in thermochemical equilibrium, focusing on what information about the planet’s mantle $f\mathrm{O}_2$ and bulk silicate H/C ratio could be determined by atmospheric observation. 800\,K volcanic atmospheres develop distinct compositional groups as the mantle $f\mathrm{O}_2$ is varied, which can be identified using sets of (often minor) indicator species: Class O, representing an oxidised mantle and containing \ce{SO2} and sulfur allotropes; Class I, formed by intermediate mantle $f\mathrm{O}_2$'s and containing \ce{CO2}, \ce{CH4}, CO and COS; and Class R, produced by reduced mantles, containing \ce{H2}, \ce{NH3} and \ce{CH4}. These atmospheric classes are robust to a wide range of bulk silicate H/C ratios. However, the H/C ratio does affect the dominant atmospheric constituent, which can vary between \ce{H2}, \ce{H2O} and \ce{CO2} once the chemical composition has stabilised to a point where it no longer changes substantially with time. This final atmospheric state is dependent on the mantle $f\mathrm{O}_2$, the H/C ratio, and time since the onset of volcanism.The atmospheric classes we present are appropriate for the closed-system growth of hot exoplanets, and may be used as a simple base for future research exploring the effects of other open-system processes on secondary volcanic atmospheres.
\end{abstract}

%% Plain language summary
%TC:ignore
\section*{Plain Language Summary}
To understand the geology of rocky planets, we must learn to use observations of their atmospheres to infer the properties of their interiors. When rocky planets have atmospheres made up of gases released by volcanoes, the chemistry of the atmosphere can be linked back to the volatile content and availability of oxygen in the planet’s interior. In order to understand what the atmospheres of these rocky planets might look like, we use computer simulations to analyse their evolution and chemical signatures. We find that planets with Venus-like surface temperatures will produce atmospheres whose chemical compositions enable them to be split into three distinct categories. These three classes are identified using diagnostic combinations of species.
%TC:endignore

\section{Introduction}

Planetary atmospheres can have three different origins: nebular gas accreted from the protoplanetary disk (primordial); early, syn-accretionary gas release from planetesimal accretion or outgassing during the cooling of a magma ocean (primary); or long-term release of volatiles from the planetary interior e.g., through volcanism (secondary). The atmospheres of rocky planets may evolve from primordial/primary to secondary as they undergo significant modification by geological and atmospheric processes, altering the atmosphere's mass fraction and chemical composition. Modification processes include hydrodynamic escape, erosion and volatile addition from impacts, and volcanic outgassing of volatiles from the interior (see Table\,\ref{table:mod_processes} for references). A more extensive list of processes which can modify planetary atmospheres, and associated literature which has investigated these effects, can be found in Table\,\ref{table:mod_processes}.

\begin{table}[ht]
\caption{Processes which can modify the chemical composition and speciation of an atmosphere, and examples of previous papers which have discussed these processes in the context of exoplanet atmospheres.}
\begin{tabularx}{\linewidth}{llX}
\toprule
\multicolumn{1}{c}{\textbf{\begin{tabular}[c]{@{}c@{}}Atmospheric\\ modification processes\\ on rocky planets\end{tabular}}} & \multicolumn{1}{c}{\textbf{\begin{tabular}[c]{@{}c@{}}Considered\\ here?\end{tabular}}} & \multicolumn{1}{c}{\textbf{Other references}}  \\ \toprule
Volcanic outgassing  & Yes  & \citeA{gaillard2014TheoreticalFrameworkVolcanic, hoolst2019ExoplanetInteriorsHabitability, liggins2020CanVolcanismBuild, ortenzi2020MantleRedoxState, kite2020ExoplanetSecondaryAtmosphere} \\ \midrule

\begin{tabular}[c]{@{}l@{}}Non-volcanic outgassing\\ (e.g., serpentinization)\end{tabular}      & No & \citeA{guzman-marmolejo2013AbioticProductionMethane}   \\ \midrule

\begin{tabular}[c]{@{}l@{}}Thermochemical kinetic\\ re-equilibration\end{tabular}               & See Paper II  & \citeA{sossi2020RedoxStateEarth, zahnle2020CreationEvolutionImpactgenerated}  \\ \midrule

Atmospheric escape   & See Paper III   & \citeA{hunten1973EscapeLightGases, walker1977EvolutionAtmosphere, lammer2014OriginLossNebulacaptured, tian2015AtmosphericEscapeSolar}  \\ \midrule

Photochemistry   & No   & \citeA{kasting2003EvolutionHabitablePlanet, hu2012PHOTOCHEMISTRYTERRESTRIALEXOPLANET, catling2017AtmosphericEvolutionInhabited, wogan2020WhenChemicalDisequilibrium, jordan2021PhotochemistryVenusLikePlanets}  \\ \midrule

\begin{tabular}[c]{@{}l@{}}Deposition, condensation\\ \& rainout\end{tabular}                   & No   & \citeA{pinto1980PhotochemicalProductionFormaldehyde, hu2012PHOTOCHEMISTRYTERRESTRIALEXOPLANET, ranjan2019NitrogenOxideConcentrations, rimmer2019OxidisedMicrometeoritesEvidence, huang2021AssessmentAmmoniaBiosignature}             \\ \midrule

\begin{tabular}[c]{@{}l@{}}Impact erosion, transformation\\ \& volatile delivery\end{tabular}   & No            & \citeA{kasting1990BolideImpactsOxidation, liuliushangfei2015GIANTIMPACTEFFICIENT, schlichting2015AtmosphericMassLoss, rimmer2019IdentifiableAcetyleneFeatures, sinclair2020EvolutionEarthAtmosphere, todd2020CometaryDeliveryHydrogen, zahnle2020CreationEvolutionImpactgenerated}    \\ \midrule

\begin{tabular}[c]{@{}l@{}}Drawdown processes (e.g., \\ silicate weathering causing\\ \ce{CO2} drawdown)\end{tabular}   & No   & \citeA{walker1981NegativeFeedbackMechanism} \\ \midrule

Biological processes  & No   & \citeA{kharecha2005CoupledAtmosphereEcosystem}  \\ \bottomrule
\end{tabularx}
\label{table:mod_processes}
\end{table}

In this paper series, we focus on secondary atmospheres which form as a result of volcanic outgassing. The chemistry of volcanic gases is dependent on the oxygen fugacity ($f\mathrm{O}_2$) of the magma \cite<and likewise the mantle from which the magma was formed, e.g.,>[]{burgisser2015SimulatingBehaviorVolatiles, gaillard2015RedoxGeodynamicsLinking, ortenzi2020MantleRedoxState}, surface pressure through volatile solubility in magmas \cite{gaillard2014TheoreticalFrameworkVolcanic}, and the relative abundance of volatile elements (i.e., H, C, O, S and N) within the magma. The secondary atmospheres of volcanically active rocky planets are therefore inextricably linked to their geological state. This is particularly useful when considering studies of exoplanets, where secondary atmospheres could provide key insight into planetary interiors through the window of volcanic activity.

The geological properties of a planet, such as size, mantle $f\mathrm{O}_2$ and volatile H/C ratio in the bulk silicate planet (the ratio of H/C in the planet's earliest mantle), are both dependent on, and in turn help to control, many facets of rocky planet evolution. The mass of a planet may control mantle $f\mathrm{O}_2$ \cite{wade2005CoreFormationOxidation}, along with the planet’s location in the protoplanetary disk and core formation processes \cite{frost2008RedoxStateMantle}. The bulk silicate H/C ratio could be dependent on the initial $^{26}$\ce{Al} content of the protoplanetary system \cite{lichtenberg2019WaterBudgetDichotomy}, degassing during early planetary differentiation \cite{hirschmann2021EarlyVolatileDepletion} and the quantity of both water and carbon in the magma ocean during core formation, which results in the sequestration of hydrogen and carbon within the core \cite{tagawa2021ExperimentalEvidenceHydrogen, grewal2021EffectCarbonConcentration}. The predicted atmospheric compositions of rocky exoplanets based on geological properties, such as mantle $f\mathrm{O}_2$ and initial volatile content, can in principle be compared to observations to test planetary formation and evolution scenarios. While properties such as the size and interior structure of a planet can already be observed, or may be inferred using mass radius observations \cite<e.g.,>[]{dornAssessingInteriorStructure2017}, establishing the mantle $f\mathrm{O}_2$ or the bulk silicate H/C ratio will require observations of atmospheric chemistry.

The composition and speciation of volcanic gases can be linked to the $f\mathrm{O}_2$ and volatile content of the magmas they originate from \cite<e.g.,>[]{holloway1987IgneousFluidsThermodynamic, gerlach1993OxygenBufferingKilauea, moretti2004OxidationStateVolatile, iacovino2015LinkingSubsurfaceSurface}. However, when degassed into an atmosphere these gases mix and are subject to various processes which will act to complicate the link between the mantle of a planet and it’s volcanic atmosphere (See Table\,\ref{table:mod_processes}). For example, as volcanic outgassing proceeds, the surface pressure increases, which in turn affects the volume and composition of the volcanic gases being emitted. At higher atmospheric pressures, water outgassing is suppressed \cite{gaillard2014TheoreticalFrameworkVolcanic}, making the emitted gases proportionally more carbon-rich. If atmospheric escape is occurring concurrently, then mass is also being lost from the atmosphere, and is acting to decrease the surface pressure. The loss of H atoms via hydrogen escape (explored in Paper III) will modify the redox chemistry of the atmosphere \cite<e.g.,>[]{kasting1993EarthEarlyAtmosphere, wordsworth2014ABIOTICOXYGENDOMINATEDATMOSPHERES}, while drawdown processes which preferentially remove certain species (e.g., the carbonate-silicate cycle removing \ce{CO2}) will affect the atmospheric composition by leaving behind elements which are less sensitive to drawdown processes  to build up in the atmosphere \protect{\cite<e.g., N;>[]{hu2019StabilityNitrogenPlanetary}}. 

Volcanic gases, once input to an atmosphere, will be cooled to the temperature of the planet's surface environment, changing the their speciation compared to that at their eruptive conditions \cite{gaillard2021DiversePlanetaryIngassing}. There will be limits on the extent of volcanic gas re-equilibration at lower atmospheric temperatures, where gases may quench at their high temperature speciations and remain in disequilibrium within the atmosphere on long timescales. Here, we focus on planets with high atmospheric temperatures, $\geq$\,800\,K, where quenching is likely to be insignificant on long timescales. Hot rocky planets are a key class of rocky exoplanet of which there is only one example in the solar system (Venus), and are the most amenable to observation in the next 5-10 years as their short period orbits increase the observational duty cycle. At 800\,K it is plausible that volcanic gasses, once added to and mixed with the atmosphere, will be able to evolve to low temperature thermochemical equilibrium. See Paper II for a complete analysis of the reactions that regulate the attainment of thermochemical equilibria in secondary volcanic atmospheres, and the temperature requirements for thermochemical equilibrium to be achieved.

Previous investigation into the formation of secondary volcanic atmospheres has focused on factors such as the mass, bulk composition, tectonic regime, early volatile content, graphite saturation of magmas, and orbital distance of a planet can affect volcanic outgassing and the resultant secondary atmosphere composition.
These past studies have identified several key features of volcanic secondary atmospheres and how they relate to the broader geodynamic state of the planet: planets which are more massive \cite<\textgreater 2$\times$ the mass of Earth,>[]{noack2017VolcanismOutgassingStagnantlid, dorn2018OutgassingStagnantlidSuperEarths}, 
which have graphite saturated magmas \cite{guimond2021LowVolcanicOutgassing}, or which have a high iron to silicon ratio \cite{spaargaren2020InfluenceBulkCompositiona} will have lower outgassing rates; stagnant lid planets may grow more massive atmospheres than those with plate tectonics \cite{spaargaren2020InfluenceBulkCompositiona} \cite<although this is controversial, e.g.,>[]{kite2009GEODYNAMICSRATEVOLCANISM, noack2014CanInteriorStructure}; and rocky planets which form with thick primordial atmospheres are less likely to have long-lived secondary atmospheres \cite{kite2020ExoplanetSecondaryAtmosphere}.
\citeA{ortenzi2020MantleRedoxState} have also recently investigated the effect of mantle $f\mathrm{O}_2$ on the atmospheres of rocky planets; however, their modelling focused on the atmospheric composition in terms of the outgassed mass of \ce{H2O}, \ce{CO2}, CO and \ce{H2} at very high temperatures (around 2000\,K) and without considering adjustment of the atmosphere to thermochemical equilibrium at surface temperatures.

Here, we model how secondary atmospheres on hot stagnant lid planets will grow and evolve over time. This regime is chosen as high surface temperatures ($>$400-600\,K) are likely to inhibit the initiation of plate tectonics, due to the lack of surface water and a reduced temperature difference between the mantle and lithosphere \cite<see>[and references within]{foley2016WholePlanetCoupling}.
The influence of mantle $f\mathrm{O}_2$, the bulk silicate H/C ratio, and atmospheric temperature is investigated, using a C-O-H-S-N volcanic outgassing model. This expands on previous work \cite<e.g.,>[]{gaillard2014TheoreticalFrameworkVolcanic, ortenzi2020MantleRedoxState} by including both sulfur and nitrogen species in the atmosphere, and accounting for the changes in speciation of gases which occur as a function of temperature. We also differ from the previous work discussed above by allowing the $f\mathrm{O}_2$ of erupted magmas (and therefore the chemistry of their associated gas phase) to evolve as they are erupted, rather than fixing them at the $f\mathrm{O}_2$ of the mantle. 

In this Paper I, we investigate the base case of volcanic secondary atmosphere formation and how these atmospheres may reflect planetary geochemistry; specifically, the mantle $f\mathrm{O}_2$ and bulk silicate H/C mass ratio. The atmosphere is solely influenced by progressive volcanic degassing, and instantaneous cooling of the atmosphere to 800\,K. These base atmospheres we present in our results are therefore a foundation upon which additional physical and chemical atmospheric processes should be tested (e.g., the kinetics of cooling atmospheres, explored in Paper II, and \ce{H2} escape in Paper III). In Sect.\,\ref{section:methods}, we present our modelling technique in full. Results are shown in Sect.\,\ref{section:results}, with discussion and conclusions presented in Sects.\,\ref{section:discuss} and \ref{section:concs}.

\section{Methods}
\label{section:methods}

\begin{figure}[h]
	\begin{center}
	 \includegraphics[width=\linewidth]{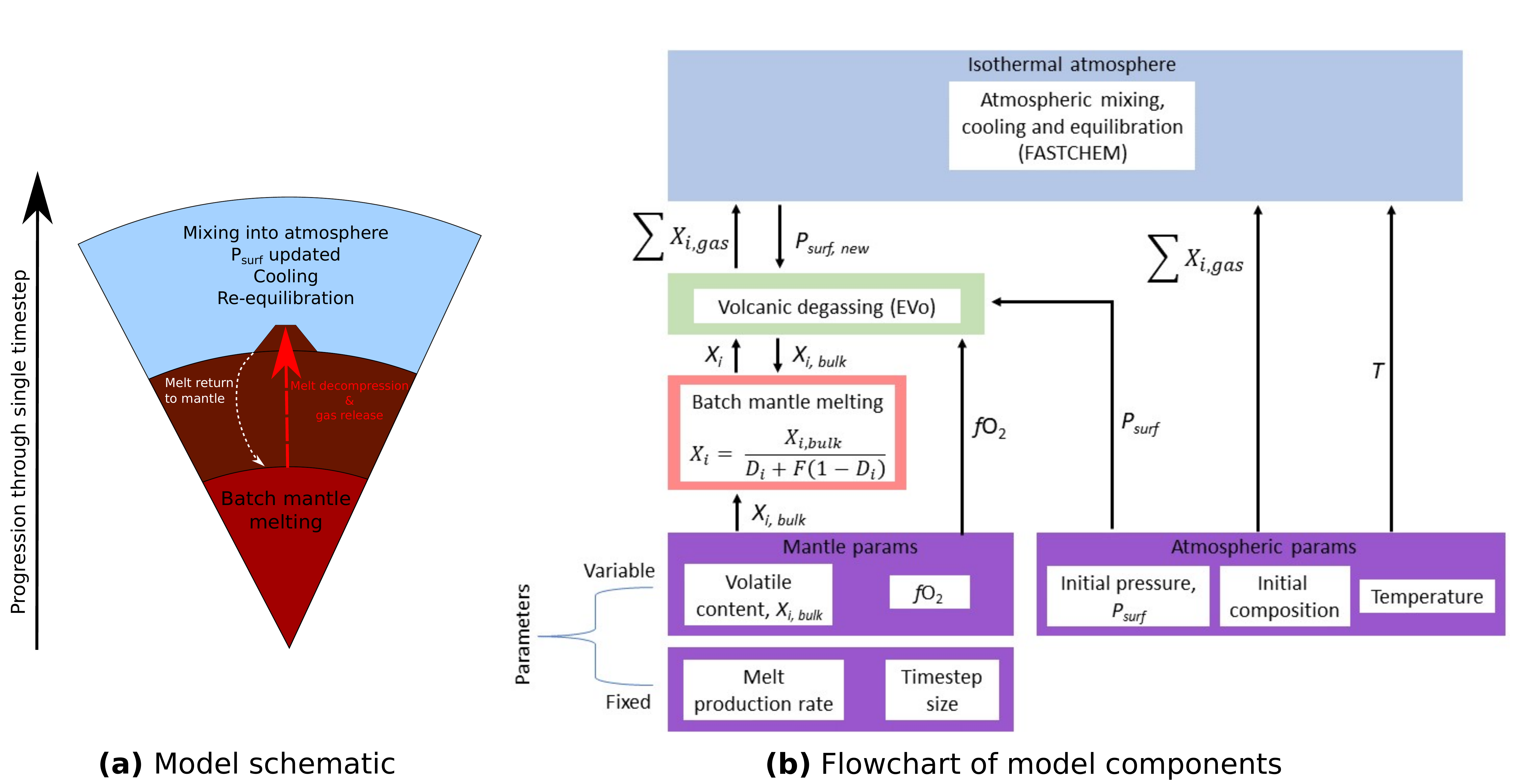}
	\caption{(a) Model schematic and (b) flowchart. Arrows in (b) show direction of information flow. See text for symbols.}
	\label{fig:model_overview}
\end{center}	
\end{figure}

To simulate the evolution of a planet's atmosphere, we have constructed EVolve, a 3-part model linking mantle (Part 1, Sect.\,\ref{section:part1}) to atmosphere (Part 3, Sect.\,\ref{section:part3}), via a volcanic system (Part 2, Sect.\,\ref{section:part2}). The relationships between each section of the model are demonstrated in Fig.\,\ref{fig:model_overview}. During a single time-step, a portion of the mantle is melted, and volatiles partition from the bulk mantle into the melt phase according to the batch melting equation (Sect.\,\ref{section:part1}). The mass and volatile content of this magma, along with the $f\mathrm{O}_2$ of the mantle, is used as an initial condition for EVo, our volcanic outgassing model, which forms Part 2 of the model (Sect.\,\ref{section:part2}). EVo returns the mass, composition, and speciation of the volcanic gas, as a mixture of 10 C-O-H-S-N species, at the current surface pressure. In the third model component (Sect.\,\ref{section:part3}), representing the atmosphere, the volcanic gas is mixed with the pre-existing atmosphere, the surface pressure is updated, and the equilibrium speciation for this atmosphere is determined using FastChem 2.0 \cite{stock2018FastChemComputerProgram, stock2021}. More detail on each of the three components is given below.

All models we present begin their calculations with 0.01 bar surface pressure (this is an arbitrary value which is small enough not to affect results, as EVo cannot be initiated with zero atmospheric pressure) from a pure \ce{N2} gas, i.e., minimal pre-existing atmosphere. This assumes that any primordial or primary atmosphere from before magma ocean solidification has been lost, or replaced by outgassing over time. Initialising a planet which retains a massive atmosphere would effectively imply the presence of a magma ocean \cite{nikolaou2019WhatFactorsAffect}, because of the strong greenhouse effect of a thick reducing atmosphere \citeA<e.g.,>[]{wordsworth2013HydrogenNitrogenGreenhouseWarming}. Such a scenario is outside the scope of this work. The escape of hydrogen to space is also neglected here, and we consider the effects of escape in detail in Paper III; this means that the results of this paper are relevant for an end-member of planetary evolution where current atmospheric escape is negligible -- an important basis on which to build models incorporating more physical/chemical processes. All models also use a single rate of melt production in the mantle. While planets are more likely to have a rate of melt production which wanes with time, as the interior of the planet cools, different rates of melt production would in practise simply change the timescale over which various points in atmospheric evolution are reached. This will not affect the interpretation of our results, as we do not aim to make detailed predictions of planets at particular points in their history.

\subsection{Part 1: Mantle Melting}
\label{section:part1}

We fix the $f\mathrm{O}_2$ of the mantle at a value relative to the iron-w{\"u}stite (IW) rock buffer, at the pressure and temperature of melt production in the mantle. We assume that melt production and the gradual loss of volatile species does not affect the $f\mathrm{O}_2$ over time, an assumption supported empirically by the relative constancy of Earth's mantle $f\mathrm{O}_2$ through time despite Earth's history of extensive volatile cycling \cite<e.g.,>[]{trail2011OxidationStateofHadeanMagmas}.
At the start of each time-step, a fraction of the mantle is melted. For an Earth-size planet, the rate of melt production in the mantle (M$_{\rm{melt}}$) is set to 1\,$\times10^{15}$\,kg\,yr$^{-1}$, based off results from \citeA{ortenzi2020MantleRedoxState}.

The volatile (H, C, S and N) content of this melt is then calculated using the batch melting equation
\begin{equation}
    X_{i,melt} = \frac{X_{i,mantle}}{D_i + F(1-D_i)},
    \label{eq:batch_melt}
\end{equation}
where $i$ indicates a single volatile species, $X_{i,melt}$ is the concentration of each volatile in the melt phase expressed as a weight fraction, $X_{i,mantle}$ is the volatile concentration in the bulk mantle, $F$ is the local melt fraction (which we fix at 0.1; we test the effect of varying the melt fraction in \ref{section:c_appendix}) and $D_i$ is the partition coefficient of the volatile (the concentration ratio of a volatile between mantle and melt at equilibrium, see Table \ref{table:partition_coeffs}). The volatile elements considered here all have partition coefficients less than 1 (C in graphite-saturated melts being the exception, discussed below), and therefore partition preferentially into the melt phase, generating a melt that is volatile element enriched in comparison to the bulk mantle.
\begin{table}[h]
\centering
\caption{Partition coefficients for bulk melting equation used in EVolve}
\begin{tabular}{lcl}
\hline
\multicolumn{1}{c}{\textbf{Species}} & \textbf{$\mathbf{D_i}$} & \multicolumn{1}{c}{\textbf{Reference}} \\ \hline
H                               & 0.01           & \cite{aubaud2004HydrogenPartitionCoefficients} \\ 
\begin{tabular}{@{}l@{}}C (graphite \\undersaturated)\end{tabular}  & 5.5\,$\times10^{-4}$ & \cite{rosenthal2015ExperimentalDeterminationPartitioning} \\
S                                      & 0.01           & \cite{callegaro2020QuintetCompletedPartitioning} \\
N                                      & \begin{tabular}{@{}c@{}}5.53\,$\times10^{-3}$ at IW, \\ 6\,$\times10^{-4}$ at NNO\end{tabular}   & \cite{li2013NitrogenSolubilityUpper}           \\
\hline
\end{tabular}
\label{table:partition_coeffs}
\end{table}

The behaviour of carbon during partial melting of the mantle is $f\mathrm{O}_2$-dependent, so a single partition coefficient does not capture it's behaviour during melting. When the mantle is sufficiently oxidised so that graphite is unstable, carbon behaves as a highly incompatible element. Under these conditions, the C content of a magma after batch melting can be calculated using the partition coefficient of \citeA{rosenthal2015ExperimentalDeterminationPartitioning} (Table\,\ref{table:partition_coeffs}). If the mantle is graphite saturated, as it will be at lower $f\mathrm{O}_2$ conditions, the amount of carbon present in the melt is controlled directly by redox equilibrium between graphite and \ce{CO3^{2-}} in the silicate melt \cite{holloway1992HighpressureFluidabsentMelting}.

Recent models of planetary degassing \protect{\cite{ortenzi2020MantleRedoxState, guimond2021LowVolcanicOutgassing}} have applied the model of \citeA{holloway1992HighpressureFluidabsentMelting} to claim that at low $f\mathrm{O}_2$, the volcanic gases feeding the atmosphere should be carbon-poor compared to more oxidised conditions, as the bulk of the planet's carbon budget remains in the mantle as graphite. However, there is a growing body of evidence which suggests that at low $f\mathrm{O}_2$, carbon can dissolve into silicate melts in forms other than \ce{CO3^{2-}} \cite<e.g.,>[]{ardia2013SolubilityCH4Synthetic, stanley2014SolubilityCOHVolatiles, armstrong2015SpeciationSolubilityReduced, dalou2019RamanSpectroscopyStudy}. We have therefore chosen to use the model of \citeA{li2017CarbonContentsReduced} under graphite saturated conditions, which calculates the total dissolved C content without speciation:

\begin{align}
    \rm{IW+1.7 \geq \log\textit{f}O_2 \geq IW-1:} \nonumber \\
    \log(\rm{C, ppm}) {}&= -3702/T - 194 P/T - 0.0034\,\log(\rm{m}_{\ce{H2O}}) \label{eq:c_high} \\&+ 0.61\,\rm{NBO/T} + 0.55\Delta \rm{IW} +3.5 \nonumber\\
    \rm{IW-5.3 \leq \log\textit{f}O_2 \leq IW-1}: \nonumber \\
    \log(\rm{C, ppm}) {}&= 0.96\,\log(\rm{m}_{\ce{H2O}}) - 0.25 \Delta \rm{IW} + 2.83 \label{eq:c_low}
\end{align}
in which $T$ is temperature in K, $P$ is pressure in GPa, m$_{\ce{H2O}}$ is the mole fraction of water in the silicate melt, $\Delta$IW is the oxygen fugacity relative to the IW buffer, and NBO/T = 2 O/T - 4 where T = Si + Ti + Al + Cr + P, the mole fractions of elements in the silicate melt \cite{li2017CarbonContentsReduced}.
We assume a mantle temperature of 2000\,K \cite<e.g., >[]{ortenzi2020MantleRedoxState} and pressure at the point of mantle melting of 2 GPa (the effect of varying the pressure of melting on our results is explored in \ref{section:c_appendix}). The mantle is assumed to no longer be graphite-saturated if the C content of the melt calculated using eq.\,\ref{eq:c_high} is higher than that calculated using the graphite-free partition coefficient of \citeA{rosenthal2015ExperimentalDeterminationPartitioning}.
If eqs.\,\ref{eq:c_high} and \ref{eq:c_low} do not intersect at precisely IW-1, we continue to use eq.\,\ref{eq:c_high} (which assumes C is dissolving as \ce{CO3^{2-}}) until the intersection is reached, this occurs within a few tenths of a log unit.

Similarly to carbon, nitrogen partitioning into a melt is also $f\mathrm{O}_2$-dependent, although it is less well studied. Two partition coefficients are provided by \citeA{li2013NitrogenSolubilityUpper}, one for the IW buffer, and one for the oxidised NNO (approx IW+4) buffer (Table\,\ref{table:partition_coeffs}). We linearly interpolate between these two values to calculate $f\mathrm{O}_2$-appropriate partition coefficients.

Following previous methods \cite<e.g.,>[]{dorn2018OutgassingStagnantlidSuperEarths, ortenzi2020MantleRedoxState}, it is assumed that 10\% of the melt produced in the mantle will reach the surface as extrusive melt, which then outgasses and contributes to the atmosphere \cite{crisp1984RatesMagmaEmplacement}. The mass of the extrusive melt which will be erupted to the surface ($M_{\rm{ext}}$, kg) is therefore calculated as
\begin{equation}
    M_{\mathrm{ext}} = \mathrm{r}_{ei}\: dt \: M_{\mathrm{melt}},
\end{equation}
where r$_{ei}$ = 0.1, the fraction of melt by mass that reaches the surface, $M_{\rm{melt}}$ (kg\,yr$^{-1}$) is the rate of melt production in the mantle and $dt$ (years) is the size of the timestep. Our rate of melt supply to the surface (1$\times10^{14}$\,kg\,yr$^{-1}$, $\sim$\,27\,km$^3$\,yr$^{-1}$ after applying r$_{ei}$ to the 1$\times10^{15}$\,kg\,yr$^{-1}$ rate of production in the mantle) falls at the lower end of estimates for melt production at mid-ocean ridges on Earth over the past 200\,Myr \cite<20 - 65\,km$^3$\,yr$^{-1}$>[]{li2016QuantifyingMeltProduction}. As emphasised above, the melt production rates only act to define the relationship between atmospheric evolution and time, rather than affect the trajectory or end point of that evolution.

Once the extrusive melt has been erupted and outgassed, calculated in Part 2 of the model (Sect.\,\ref{section:part2}), the extrusive melt and any residual volatiles that did not get released into the atmosphere are mixed back into the mantle reservoir, along with the intrusive melt. The new volatile content of the mantle after each outgassing time-step is calculated by recombining the three reservoirs, by simple mass balance

\begin{equation}
    X_{i,mantle}(t+dt) = \frac{X_{i,ext}(t)\: M_{\rm{ext}}(t) + X_{i,int}(t)\:M_{\rm{int}}(t) + X_{i,mantle}(t)\:M_{mantle}(t)}{M_{\rm{ext}}(t)+M_{\rm{int}}(t)+M_{mantle}(t)}.
\label{eq:mantle_mixing}
\end{equation}
where $M$ is the mass of each reservoir, and the subscripts $int$ and $ext$ refer to the intrusive and extrusive magmas, respectively. The lithospheric mass remains constant over time, as all the silicate melt is returned to the mantle after each time-step. This approach is conceptually consistent with a planet operating in a stagnant-lid regime, equivalent to assuming a slow return of crustal material and it's embedded volatiles to the convecting mantle via delamination or lithospheric `drip' \cite{stern2018StagnantLidTectonics}. Atmosphere-interior volatile cycling is assumed to be inefficient, i.e., volatiles remain in the atmosphere once outgassed \cite{tosi2017HabitabilityStagnantlidEarth}. This is appropriate for the hot planets we consider here, on which low temperature, aqueous drawdown processes such as the carbonate-silicate cycle do not operate \cite<including those suggested by>[for stagnant lid planets]{foley2019HabitabilityEarthlikeStagnant, honing2019CarbonCyclingInterior}.

\subsection{Part 2: EVo, Volcanic Outgassing}
\label{section:part2}

To calculate the gas composition input to an atmosphere from a magmatic source, we use a model employing the `equilibrium constant and mass balance' method \cite<see>[]{holloway1987IgneousFluidsThermodynamic}, built on the EVo code previously described in \citeA{liggins2020CanVolcanismBuild}. This model calculates the speciation and volume of a C-O-H-S-N gas phase (composed of 10 species: \ce{H2O}, \ce{H2}, \ce{O2}, \ce{CO2}, CO, \ce{CH4}, \ce{S2}, \ce{SO2}, \ce{H2S} and \ce{N2}) in equilibrium with a silicate melt at a given pressure, temperature and magma $f\mathrm{O}_2$, considering both homogeneous gas-phase equilibria described in eqs.\,\ref{eqn:k1}-\ref{eqn:k5} 

\begin{align}
	\ce{H2 + \frac{1}{2} O2 &<=> H2O}, \label{eqn:k1}\\
	\ce{CO + \frac{1}{2} O2 &<=> CO2}, \label{eqn:k2}\\
	\ce{CH4 + 2O2 &<=> CO2 + 2H2O}, \label{eqn:k3}\\
	\ce{H2S + \frac{1}{2} O2 &<=> \frac{1}{2} S2 + H2O}, \label{eqn:k4}\\
	\frac{1}{2}\ce{S2 + O2 &<=> SO2}, \label{eqn:k5}
\end{align}
and heterogeneous gas-melt equilibria considered in the form of solubility laws suitable for a melt with a basaltic composition. These solubility laws are taken from \citeA{burgisser2015SimulatingBehaviorVolatiles} for \ce{H2O} and \ce{H2}, \citeA{eguchi2018CO2SolubilityModel} for \ce{CO2}, \citeA{armstrong2015SpeciationSolubilityReduced} for CO, \citeA{ardia2013SolubilityCH4Synthetic} for \ce{CH4} (both CO and \ce{CH4} are set to insoluble if the starting $f\mathrm{O}_2$ is above IW+1), \citeA{libourel2003NitrogenSolubilityBasaltic} for \ce{N2}, and using the sulfide capacity law of \citeA{oneill2020ThermodynamicControlsSulfide} to calculate \ce{S2-} solubility as is appropriate for sulfur in reduced melts. Oxygen exchange between the gas phase and iron in the melt is prescribed according to \citeA{kress1991CompressibilitySilicateLiquids} as
\begin{equation}
    \ce{Fe2O3_{(melt)} <=> 2FeO_{(melt)} + 0.5O2_{(gas)}},
    \label{eq:iron_buffer}
\end{equation}
so the gas is always in $f\mathrm{O}_2$ equilibrium with the silicate melt. While the $f\mathrm{O}_2$ of the system is allowed to evolve away from the initial mantle $f\mathrm{O}_2$ during decompression and outgassing, the iron pool in the melt acts as a buffer to this change \cite{burgisser2007RedoxEvolutionDegassing}.

During a single time-step, EVo is initialised to first find the volatile saturation pressure based on the $f\mathrm{O}_2$ and weight fractions of C, H, S and N in the melt. An iterative solver from the SciPy package of \citeA{virtanen2020SciPyFundamentalAlgorithms} is used to calculate both the distribution of CHSN elements between the different species in the melt (e.g., C across \ce{CO3-}, \ce{CO2}, CO and \ce{CH4}) according to the mantle $f\mathrm{O}_2$, and the corresponding volatile saturation pressure. When the saturation pressure of a magma with a given dissolved volatile concentration is found, the following equality holds:
\begin{equation}
    \sum_{i=1} P_i - P = 0
\end{equation}
where $P$ is the total pressure and $P_i$ is the partial pressure of species $i$ calculated according to it's corresponding solubility law, for a fixed concentration in the melt.

Once this starting pressure has been found, EVo calculates the outgassing path of the system to the surface pressure, assuming a constant temperature of 1200\textdegree{}C/1473\,K. As the system pressure is lowered step-wise, the homogeneous gas equilibrium equations for reactions \ref{eqn:k1} - \ref{eqn:k5}, gas-melt equilibria and eq.\,\ref{eq:iron_buffer} are solved simultaneously, conserving the total mass of each volatile element across the system (gas + melt). The system is assumed to have reached thermochemical equilibrium at every pressure step.

\subsection{Part 3: Atmospheric Processing}
\label{section:part3}

Atmospheric processing in EVolve is calculated using FastChem 2.0 \cite{stock2018FastChemComputerProgram, stock2021}.
Once the mass and chemistry of a volcanic gas input to the atmosphere during a time-step has been calculated in Part 2 (using EVo), the total element abundances of the new, mixed atmosphere (pre-existing atmosphere + volcanic gas) are provided to FastChem. The new atmospheric pressure is calculated as 
\begin{equation}
    P_{\mathrm{j}} = P_{\mathrm{j-1}} + \frac{M_{\mathrm{ext}} \,W_{\rm{g}} \, g}{4 \pi R_{\rm{P}}^2},
    \label{eq:pressure}
\end{equation}
where $P_{\rm{j}}$ (Pa) is the surface pressure after the release of volcanic gases, $P_{\rm{j-1}}$ is the surface pressure produced by the pre-existing atmosphere from the previous time-step (the pressure at which the volcanic gas is erupted at), $W_{\rm{g}}$ is the weight fraction of exsolved gas in the volcanic system, $g$ (m s$^{-2}$) is the surface gravity and $R_{\rm{P}}$ (m) is the radius of the planet.

The final atmospheric speciation, in thermochemical equilibrium at the pre-defined atmospheric temperature, is calculated at this new surface pressure. The atmospheric chemistry is calculated as a function of all 85 species considered by FastChem, and so includes many species beyond those listed in eqs.\,\ref{eqn:k1} - \ref{eqn:k5}. The surface temperature is fixed, irrespective of atmospheric chemistry; this removes the effect of an evolving climate on our model results. Calculating the surface temperature self-consistently using the evolving atmospheric chemistry goes beyond the scope of this paper, but should be explored in future work to evaluate its effects on our results. This is a reasonable simplification, as given that the atmosphere is assumed to be in instantaneous thermochemical equilibrium, the surface temperature history does not affect the final atmospheric chemistry at any point. For simplicity, and due to a lack of representative P-T profiles for the atmosphere as it evolves, the atmospheric speciation is only calculated at surface pressure. Hydrogen escape has also been included in the atmospheric component of EVolve, but we leave discussions of this to Paper III.

\section{Results}
\label{section:results}

We investigate the effects of mantle $f\mathrm{O}_2$ and atmospheric temperature on the atmospheric growth and evolution of Earth-sized, stagnant lid planets. A single starting mantle volatile content is initially considered, based on the \ce{H2O} and \ce{CO2} content of the mantle after magma ocean solidification in a volatile-rich delivery scenario \cite<450\,ppm \ce{H2O} and 50\,ppm \ce{CO2};>[]{elkins-tantonLinkedMagmaOcean2008}. The mantle volatile contents of exoplanets are highly uncertain, and in the absence of any other reliable data on the S and N contents of planetary mantles the initial concentrations of sulfur and nitrogen in the model mantles are set by scaling estimates of Earth's depleted MORB (mid-ocean ridge basalt) source mantle content (\citeA<140\,ppm \ce{CO2}:>{levoyer2017HeterogeneityMantleCarbon}; \citeA<150\,ppm S:>{ding2017FateSulfideDecompression}; \citeA<1\,ppm N:>{marty2003NitrogenRecordCrust}), to match the \ce{CO2} value from \citeA{elkins-tantonLinkedMagmaOcean2008}. A single rate of mantle melting is considered, so all times since the onset of volcanism discussed here are linearly dependent on the melt production rate -- i.e., time and cumulative eruptive melt volume are interchangeable, and our use of time to track atmospheric evolution is simply indicative (see Fig.\,\ref{fig:through_time} for examples of melt volumes implied by a given age). All planets are assumed to be Earth-sized, and are initialised with a 0.01 bar \ce{N2} atmosphere. Atmospheric chemical compositions are presented at surface pressure.

We explore the effect of atmospheric temperature on volcanic atmospheres at thermochemical equilibrium (Sect.\,\ref{section:temp-effects}); how the resulting speciation of the lower temperature volcanic atmospheres fall into distinct atmospheric classes depending on their mantle $f\mathrm{O}_2$ (Sect.\,\ref{section:fo2-classes}); the effect of the bulk silicate H/C ratio on these atmospheric classes (Sect.\,\ref{section:hc_ratio}) and how volcanic atmospheres evolve over long timescales (Sect.\,\ref{section:long-term}).

\subsection{Volcanic Atmospheres in Thermochemical Equilibrium}
\label{section:temp-effects}

We first compare results where volcanic gases have reached thermochemical equilibrium at three different planetary surface temperatures: 2000, 1500 and 800\,K.

\begin{figure}[h]
    \centering
    \includegraphics[width=0.75\textwidth]{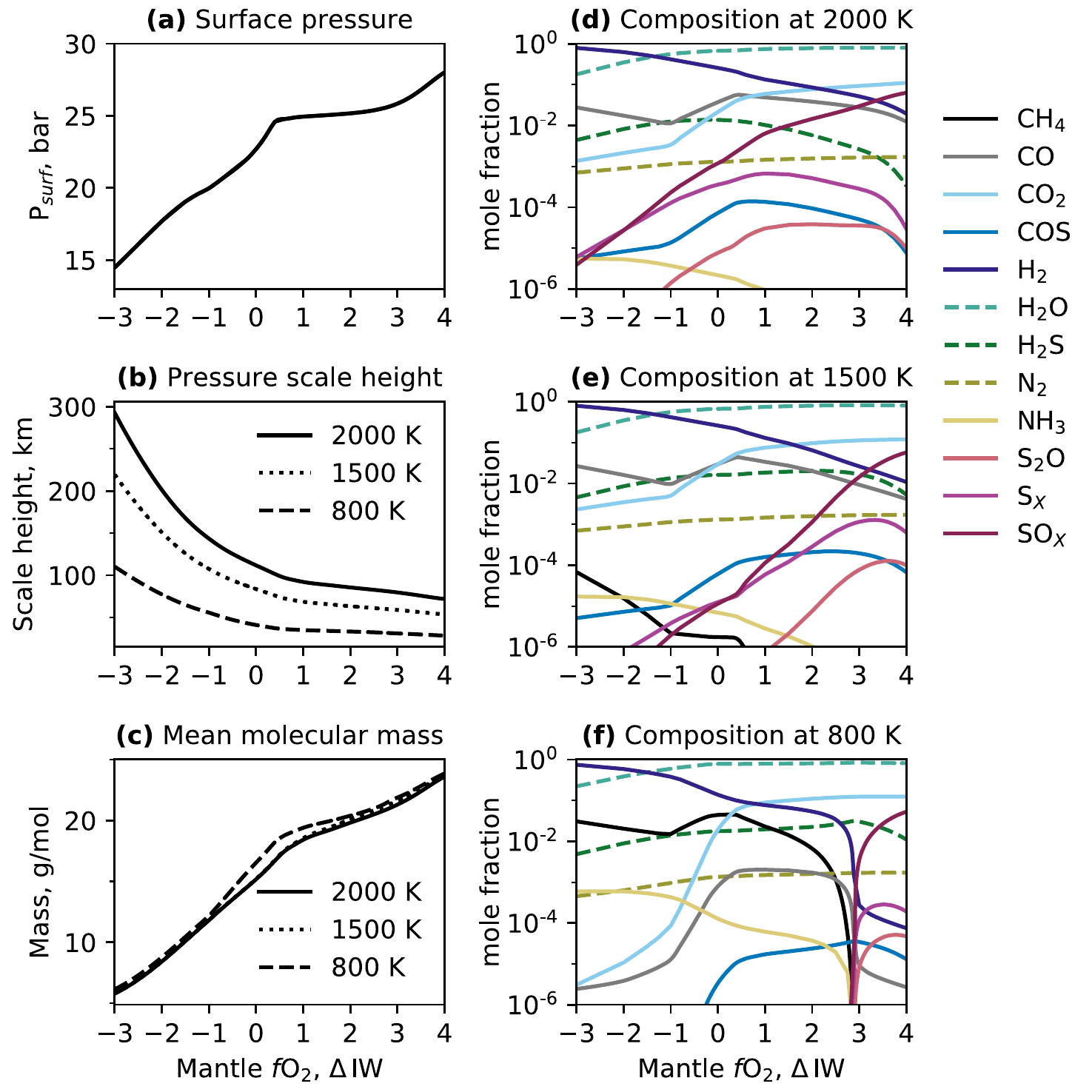}
    \caption{The effect of isothermal atmospheric temperature on the physical properties: (a) surface pressure, (b) scale height, and (c) mean molecular mass; and atmospheric chemistry at surface pressure: (d) at 2000\,K, (e), at 1500\,K, and (f) at 800\,K, after 1\,Gyr of volcanic activity. Atmospheres $\geq$\,1500\,K have greater scale heights and a more gradual change in atmospheric chemistry with mantle $f\mathrm{O}_2$ compared to 800\,K isothermal atmospheres. Ions (notably \ce{OH-}, \ce{HS-} and \ce{O^{2-}}) have been excluded from the 2000\,K chemistry plot, but are present at approximately constant levels across the $f\mathrm{O}_2$ range presented here. The full data results for these plots, including ions, are available in the data repository.}
    \label{fig:no_escape}
\end{figure}

The surface pressure, scale height, mean molecular mass, and major atmospheric chemistry for three different surface temperature cases are shown in Fig.\,\ref{fig:no_escape}, as functions of mantle $f\mathrm{O}_2$. Results are shown after 1\,Gyr of volcanic activity.
Figure\,\ref{fig:no_escape}(a) shows that the surface pressure, while unaffected by temperature, increases non-linearly with mantle $f\mathrm{O}_2$.
The increase in surface pressure with mantle $f\mathrm{O}_2$ is controlled by a number of factors: (1) a more oxidised mantle will produce more oxygen-bearing gas species, with a correspondingly higher mean molecular weight (e.g., \mbox{\ce{CO2}} replacing CO). (2) The partitioning of both C and N into the melt phase during mantle melting is $f\mathrm{O}_2$ dependent, where a smaller fraction of the volatile budget enters the melt at low $f\mathrm{O}_2$ (see Sect.\,\mbox{\ref{section:part1}}). (3) The $f\mathrm{O}_2$ dependence of many gas-melt solubility laws; at low $f\mathrm{O}_2$, nitrogen outgassing from the melt phase is suppressed \cite<it becomes much more soluble by speciating into the melt as \ce{N^{3-}}>[]{libourel2003NitrogenSolubilityBasaltic}, while at high $f\mathrm{O}_2$ sulfur outgassing is enhanced \cite<the sulphide capacity of the melt lowers under more oxidising conditions, e.g.,>[]{oneill2020ThermodynamicControlsSulfide}. This creates the stepped trend in surface pressure, with more massive atmospheres formed at higher mantle $f\mathrm{O}_2$.

Conversely, the atmospheric scale height decreases with increasing mantle $f\mathrm{O}_2$, reflecting the lower mean molecular weight of atmospheres on planets with reduced mantles, and particularly their increased abundance of \ce{H2} \cite{miller-ricci2008ATMOSPHERICSIGNATURESSUPEREARTHS}. The effect of atmospheric temperature on the mean molecular weight of the atmosphere is small (maximum 3\,g/mol for a given $f\mathrm{O}_2$, Fig.\,\ref{fig:no_escape}(c)), and therefore the differences in scale height between the three atmospheres are almost entirely due to thermal expansion with the 2000\,K atmosphere having a scale height $\approx$ 2.5$\times$ greater than the 800\,K atmosphere across the $f\mathrm{O}_2$ range.

Figures\,\ref{fig:no_escape}\,d-f demonstrate that the atmospheric chemistry of volcanic atmospheres changes systematically as a function of $f\mathrm{O}_2$ and surface temperature. The atmospheric chemistry of the 1500\,K atmosphere agrees with trends shown in previous work \cite<e.g.,>[]{ortenzi2020MantleRedoxState}; there is a smooth transition from the atmosphere being rich in \ce{H2O}, \ce{CO2} and \ce{SO$_x$} (in this case speciated almost entirely as \ce{SO2}) under oxidised mantle scenarios, to being more rich in \ce{H2} and CO under reduced scenarios, and with \ce{CH4} abundance increasing as mantle $f\mathrm{O}_2$ reduces below IW+1.
The atmospheric chemistry of the 2000\,K atmosphere shows less variability with mantle $f\mathrm{O}_2$ than that of the 1500\,K atmosphere; \ce{CH4} is absent at the ppm level across the entire $f\mathrm{O}_2$ range, while \ce{SO$_x$} species are present even at the most reduced mantle $f\mathrm{O}_2$ conditions.
Ions are also present in the 2000\,K atmosphere (including \ce{OH-} and \ce{HS-}). As none of the ions show a strong change in abundance with mantle $f\mathrm{O}_2$, they have been omitted from Fig.\,\ref{fig:no_escape} for clarity of results.

In contrast, the atmospheric chemistry of the 800\,K atmosphere shows much more variability with mantle $f\mathrm{O}_2$, compared with both the 2000\,K and the 1500\,K atmospheres. Rather than a smooth transition from oxidising to reducing atmospheres, the atmospheric chemistry shows distinct transitions in the abundance of certain species, which coincide with inflections in the total atmospheric pressure (Fig.\,\ref{fig:no_escape}a). A sharp transition in the atmospheric speciation can be seen at high $f\mathrm{O}_2$ (IW+3), where oxidised sulfur species (\ce{SO$_x$}, \ce{S2O}, S$_x$) drop below ppm abundances, and the \ce{CH4} content rapidly increases to $\sim$\,1\%. Another, more gradual transition can be seen at low $f\mathrm{O}_2$ (IW), where the \ce{CO2} and CO abundances drop sharply to less than 10\,ppm, COS decreases to below ppm levels and the \ce{H2} content increases again to $>$\,10\%. At this point, the \ce{NH3} abundance also exceeds CO, reaching $>$\,500\,ppm at a mantle $f\mathrm{O}_2$ of IW-2.

\subsection{Atmospheric classes on hot planets}
\label{section:fo2-classes}

The results of Section\,\ref{section:temp-effects} have shown that once volcanic atmospheres cool to below eruptive temperatures, they start to form more distinct compositional groups, linked to the $f\mathrm{O}_2$ of the mantle supplying the degassing magmas (Fig.\,\ref{fig:no_escape}). Here we explore how these compositional groups can be classified in an 800\,K atmosphere, after 1\,Gyr of outgassing (at which point these atmospheric compositions become stable in time).

\begin{figure}[h]
    \centering
    \includegraphics[width=0.5\textwidth]{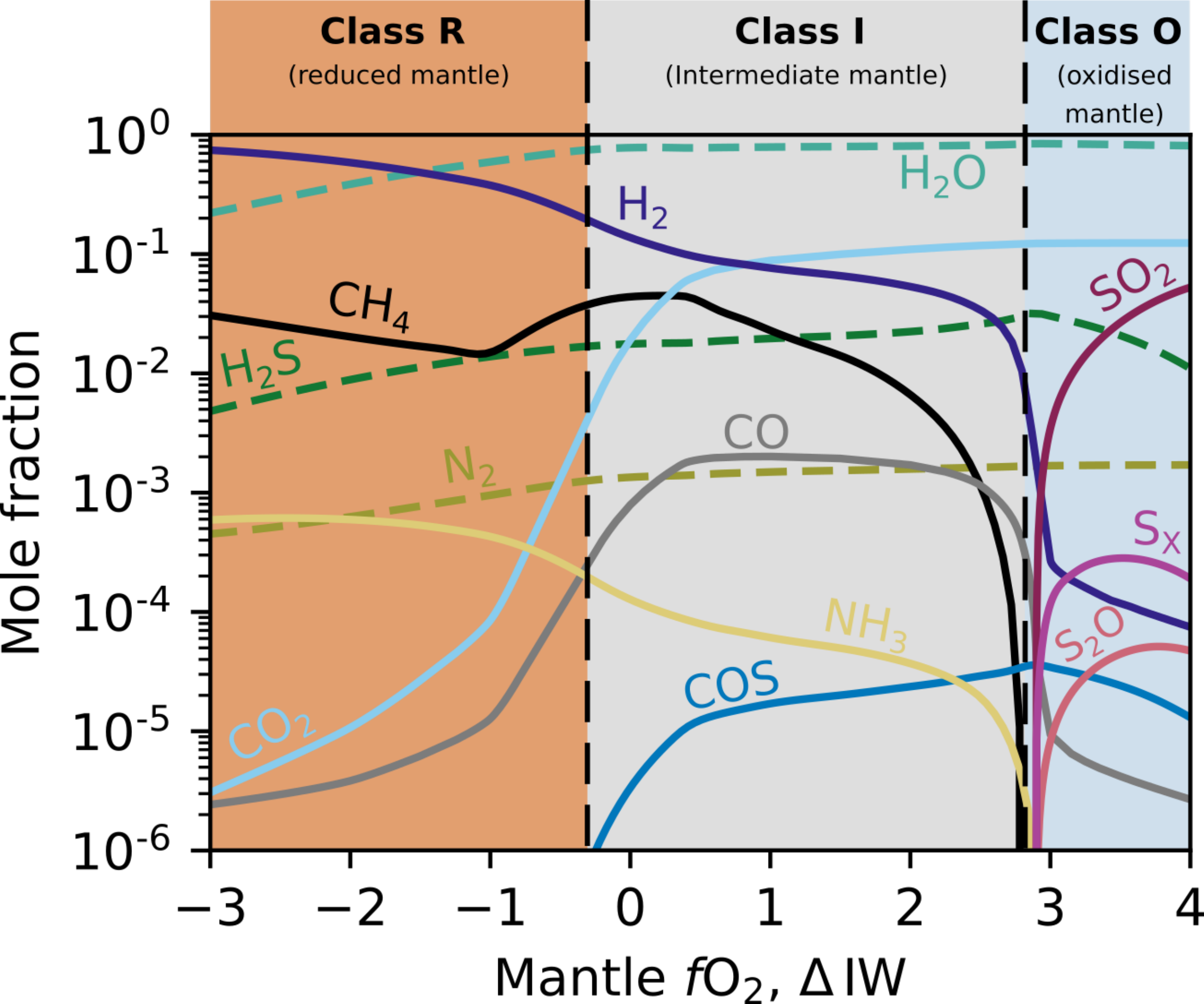}
    \caption{The three secondary volcanic atmosphere classes. Each atmospheric class is indicative of the underlying mantle $f\mathrm{O}_2$, with each distinguished by the changing abundance of key species: \ce{SO2}, \ce{S2O} and S$_x$ for Class O; \ce{CH4}, \ce{CO2}, CO and COS for Class I; and \ce{H2}, \ce{CH4} and \ce{NH3} for class R. The species irrelevant to the class definitions (\ce{H2O}, \ce{H2S}, \ce{N2}) have been plotted in dashed lines. Compositions shown for an 800\,K atmosphere after 1\,Gyr of volcanic outgassing.}
    \label{fig:comp_groups}
\end{figure}

By considering the species which vary by more than an order of magnitude in abundance across the mantle $f\mathrm{O}_2$ range for the 800\,K atmosphere, three classes of secondary volcanic atmospheres can be defined (Fig.\,\ref{fig:comp_groups}): 

\emph{Class R} atmospheres, present on planets with reduced mantles ($<$IW-0.5), are defined by the presence of \ce{H2} and \ce{CH4} with mixing ratios $>$1\%, alongside \ce{NH3} and very low or declining levels of \ce{CO2} and CO. Class R atmospheres are also more extended, with significant \ce{H2} inflating the scale height.

\emph{Class I} atmospheres, produced by planets with intermediate mantle $f\mathrm{O}_2$ between approximately IW-0.5 and IW+2.7, are characterised by the presence of \ce{CO2}, \ce{CH4} at mixing ratios $<$1\%, alongside smaller amounts of CO and COS.

\emph{Class O} atmospheres, are formed by planets with oxidised mantles ($>$IW+2.7), and are classified by the presence of \ce{SO2} and sulfur allotropes (S$_x$).

\ce{H2O}, \ce{H2S} and \ce{N2} are present across the entire $f\mathrm{O}_2$ range here and are found in all three atmospheric types. These three classes show the chemistry of an atmosphere can be directly linked to the mantle $f\mathrm{O}_2$ of a planet, even after volcanic gases are allowed to react in the atmosphere and cool down from their eruptive temperatures -- subject to no further modification processes such as escape.

\begin{figure}[h]
    \centering
    \includegraphics[width=0.5\textwidth]{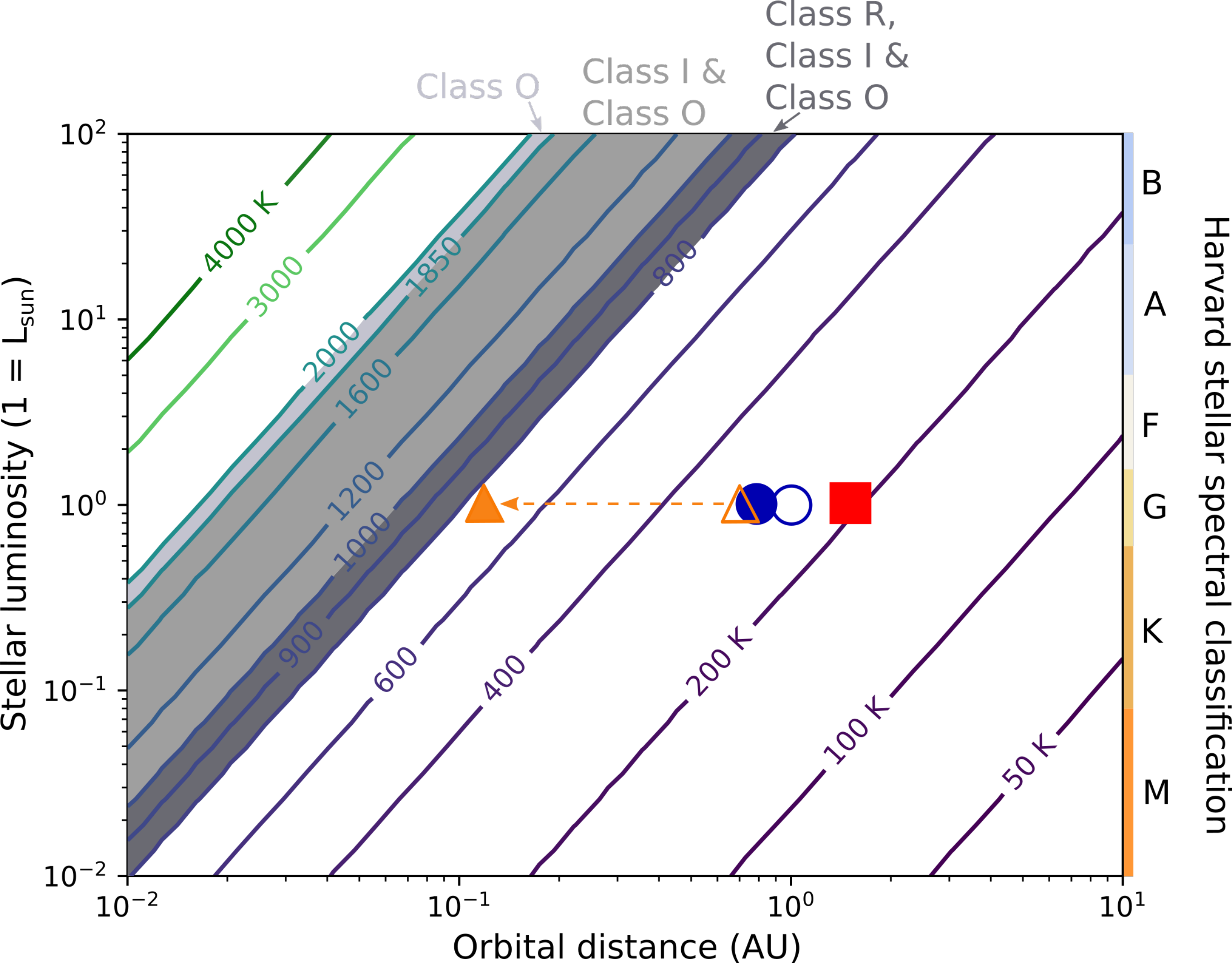}
    \caption{Equilibrium temperature of planets assuming an Earth-like Bond albedo, with atmospheric classes possible (over an IW-3 to IW+4 mantle $f\mathrm{O}_2$ range) according to the luminosity of the star with respect to the Sun (a property controlled by their effective temperature and radius), and the orbital distance of the planet superimposed. Temperature contours are for the equilibrium temperature of a planet in kelvin. Earth (blue circle) and Venus (orange triangle) and Mars (red square) are plotted at their orbital distances, indicating their equilibrium temperature if they had a Bond-albedo similar to Earth's (empty symbol) and their actual surface temperatures (filled symbol) as controlled by the climatic conditions created by their atmospheres. The formation history, and therefore initial volatile content of planets is assumed to be constant, regardless of the planet's distance from the star.}
    \label{fig:stellar_lines}
\end{figure}

As shown in Figs.\,\ref{fig:no_escape}\,d-f, the $f\mathrm{O}_2$ dependence of atmospheric speciation diminishes at higher temperatures, and not all of the three atmospheric classes defined above (for an 800\,K atmosphere) will be present for planets with higher atmospheric temperatures. Figure\,\ref{fig:stellar_lines} shows the approximate equilibrium temperature of a planet according to the luminosity of it's parent star (a function of stellar radius and temperature), and it's orbital distance in AU. Superimposed on top are the atmospheric classes which could be present (over the mantle $f\mathrm{O}_2$ range IW-3 to IW+4), assuming the atmospheric temperature is equal to the equilibrium temperature. Cooler planets can exhibit the characteristics of a Class R atmospheres if the mantle $f\mathrm{O}_2$ is sufficiently low, however once the atmosphere is hotter than $\sim$\,1850\,K the atmospheric chemistry will resemble a Class O atmosphere, regardless of how reduced the planetary mantle is (see \ref{section:t_appendix} for plots showing these T-dependent class changes). Above 2000\,K, atmospheres start to contain a significant abundance of ions not considered here, so no longer fit into the Class O classification well. Discussions of atmospheres below 800\,K are left to Paper ii.

The atmospheric temperature of a planet may also be warmer than the equilibrium temperature plotted here, depending on the composition and thickness of it's atmosphere. As an example of this, the greenhouse effect of the present atmospheres of Venus, Earth and Mars are shown in Fig.\,\ref{fig:stellar_lines} with triangles, circles and squares, respectively. The empty symbols indicate each planet's equilibrium temperature based on their orbital distances from the Sun, while the filled symbols indicate the true surface temperature of each planet that results from atmospheric greenhouse warming. Planets further out from their stars may therefore also fall into the temperature bands shown here where our classification system applies, depending on their climate.

\subsection{Effect of the Bulk Silicate H/C Ratio}
\label{section:hc_ratio}

\begin{figure}[h!]
  \begin{minipage}[l]{0.5\textwidth}
	  \includegraphics[width=\linewidth]{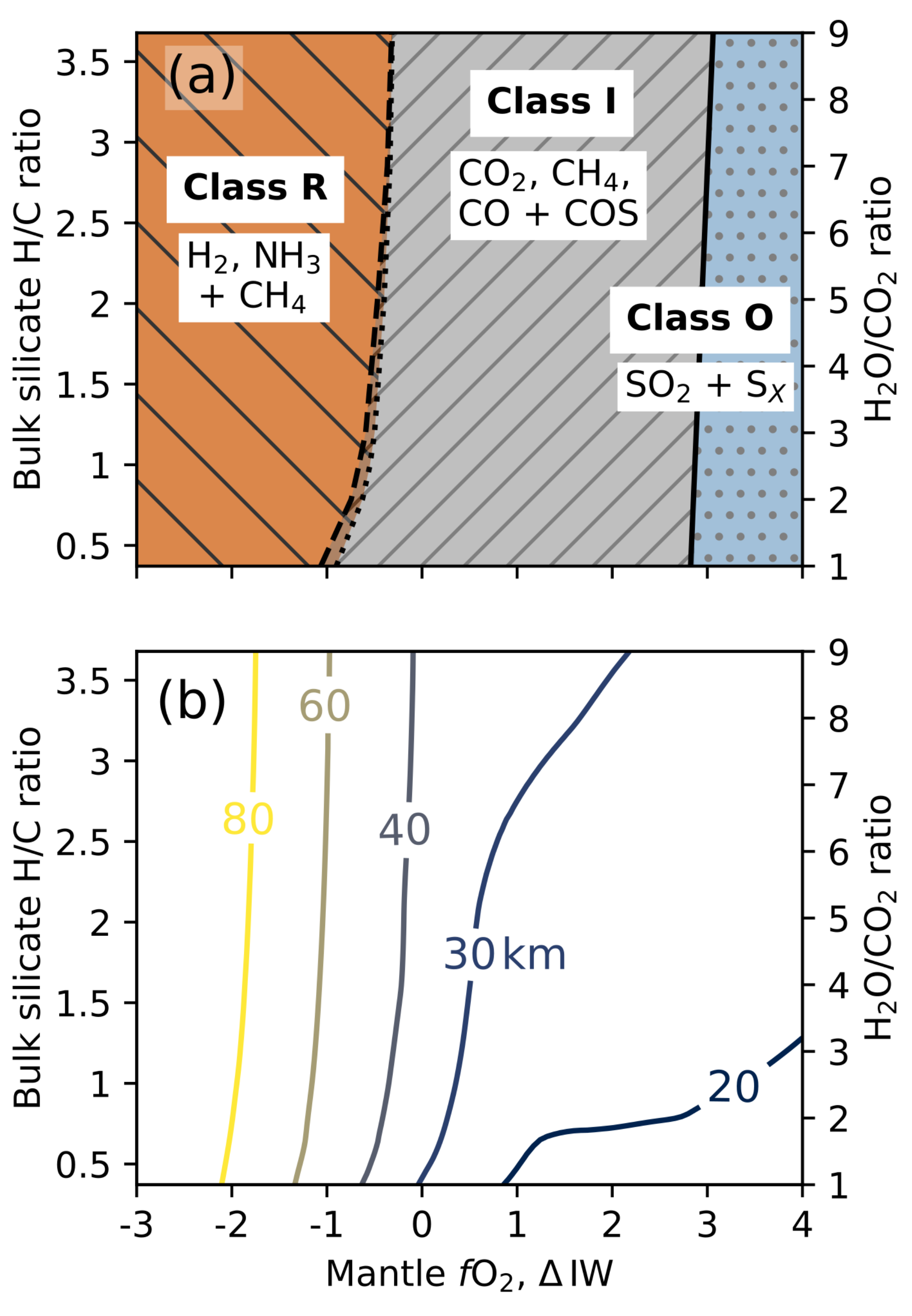}
  \end{minipage}\hfill
  \begin{minipage}[c]{0.45\textwidth}
    \caption{(a) Volcanic secondary atmosphere regime diagram. The three identified secondary volcanic atmosphere classes on mature planets at 800\,K are shown, varying with mantle $f\mathrm{O}_2$ and initial H/C ratio, after 3\,Gyr of outgassing. Two dashed lines indicate the two different markers for a class R atmosphere; the dotted line indicates the point where \ce{NH3} becomes more abundant than CO, the dashed where the COS abundance drops below 1\,$\times10^{-6}$ (these lines fall very close to each other). The bulk silicate H/C ratio is also presented as its equivalent \ce{H2O}/\ce{CO2} mass ratio on the right-hand axis, assuming all hydrogen in the mantle is speciated as \ce{H2O}, and all carbon as \ce{CO2}. (b) Pressure scale height (km) at 3\,Gyr with mantle $f\mathrm{O}_2$ and bulk silicate H/C ratio.}
    \label{fig:hc_ratio}
  \end{minipage}
\end{figure}

The ratio of different volatiles in a planet's mantle can also affect the chemistry of volcanic gases released by making them more or less carbon rich, for example. The robustness of the atmospheric classifications listed above (calculated using a bulk silicate H/C ratio of approximately 3.5) to a variable bulk silicate H/C mass ratio is tested in Fig.\,\ref{fig:hc_ratio}. The ratio of C:S:N remains fixed, so the only variable changing is the proportional hydrogen content. We present results for atmospheres after 3\,Gyr of volcanic activity, because in some simulations with a mantle $f\mathrm{O}_2$ close to class transitions and a low H/C ratio (water-poor), it took longer than 1\,Gyr to reach a class composition. In each case after 3\,Gyr the class transitions were then maintained out to 10\,Gyr. This effect of a compositional dependence on the timescale of atmospheric evolution, and the few edge cases which remain where the atmospheric class continues to change after 3\,Gyr, are explored further in Sect.\,\ref{section:long-term}.

Class O atmospheres (\ce{CH4} $<$\,1\,$\times10^{-6}$, high \ce{SO2} and \ce{S_X}) appear for a mantle $f\mathrm{O}_2$ more oxidised than $\sim$\,IW+3 across a wide range of bulk silicate H/C ratios.
The transition from Class R to I is also largely constant at IW-0.5, although at H/C$=$1 and below this occurs at slightly more reducing conditions, towards IW-1. Fig.\,\ref{fig:hc_ratio} shows that while mantle $f\mathrm{O}_2$ may be classified using the atmospheric speciation, the same cannot be said for the bulk silicate H/C ratio.
Over the H/C range shown here, there are no distinct changes in either the speciation of the atmosphere, or the pressure scale height which might indicate the initial H/C ratio. Pressure scale heights (directly proportional to atmospheric extent) of $\geq$ 60\,km indicates a class R atmosphere for an 800\,K isothermal atmosphere.

\subsection{Long-Term Atmospheric Evolution}
\label{section:long-term}

\begin{figure}[h!]
	\begin{center}
	\includegraphics[width=\linewidth]{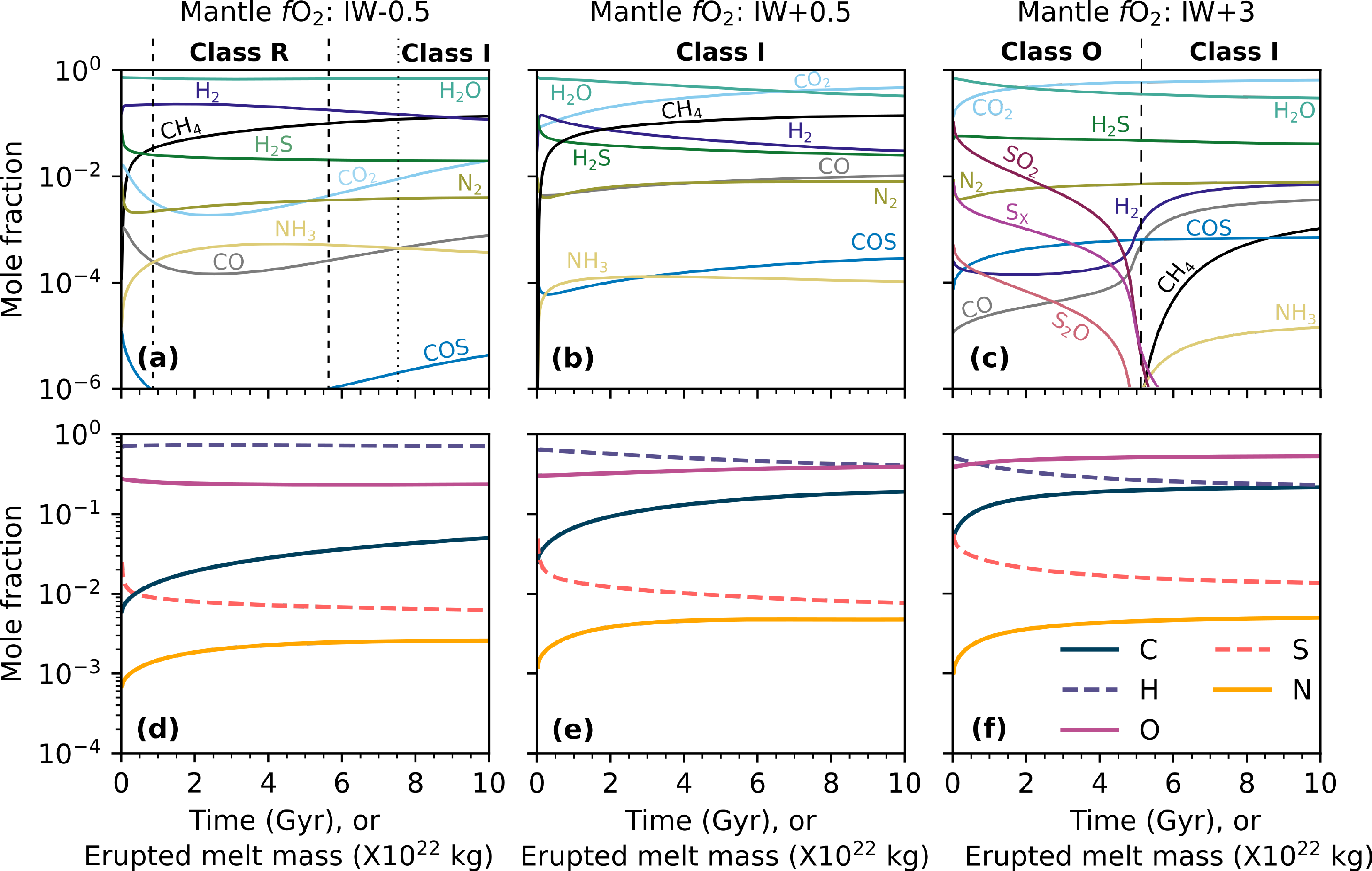}
	\caption{The evolution of three different 800\,K atmospheres over 10\,Gyr of time, showing the speciation of the atmosphere (a-c) and the atmospheric composition as mole fractions of each atomic element (d-f). (a) and (c) have a H/C=2, (b), (c) (e) and (f) have H/C=1. Most scenarios follow the pattern of (b), stabilising into a single compositional class after 1-3\,Gyr. Models which sit at the edge of classes, such as those in (a) and (c), may change groups over longer time periods.}
	\label{fig:through_time}
\end{center}	
\end{figure}

\begin{figure}[h!]
    \centering
    \includegraphics[width=0.5\textwidth]{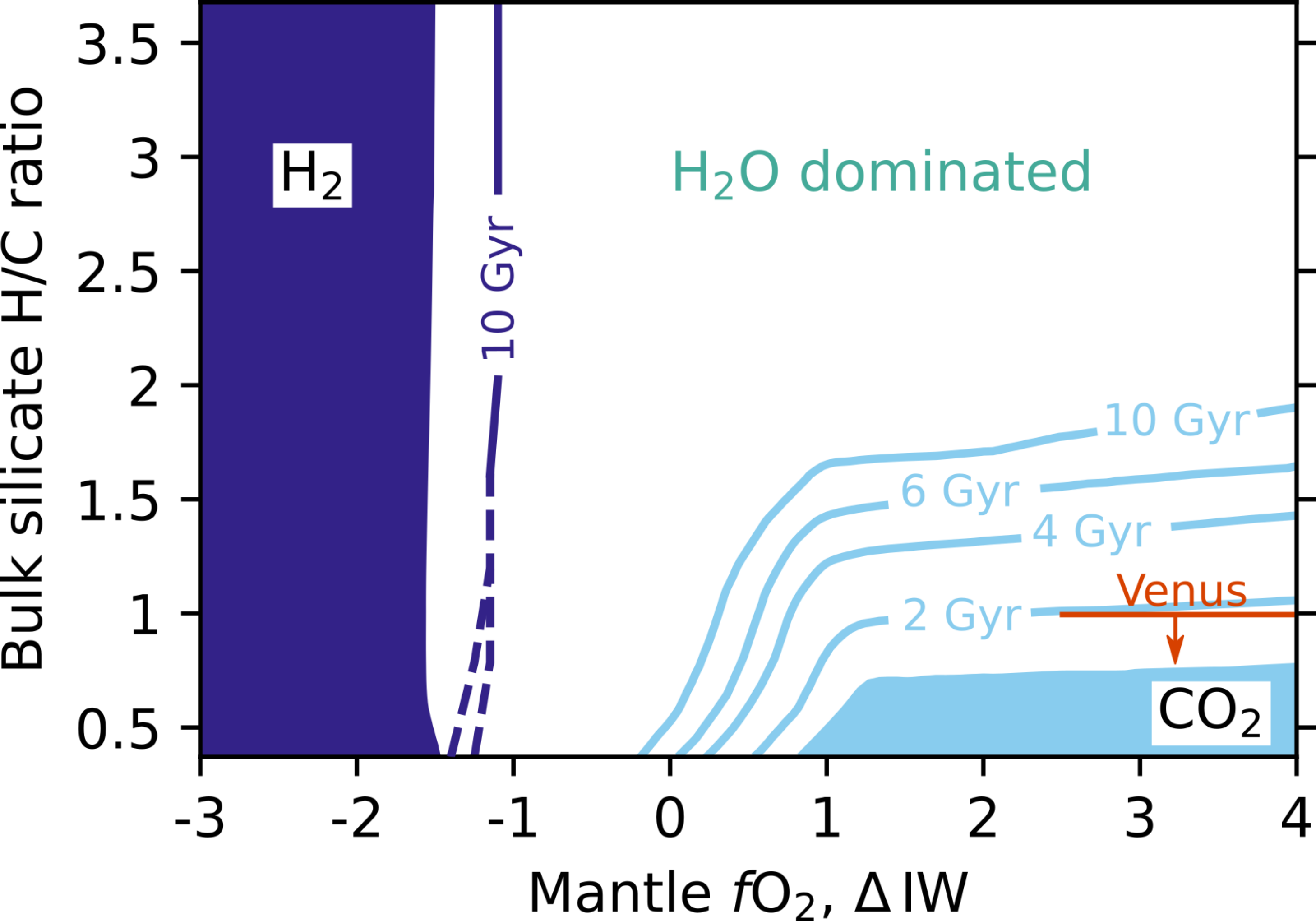}
    \caption{The time taken (in Gyr) for a planetary atmosphere to become dominated by a species other than \ce{H2O}, for a given bulk silicate H/C ratio and mantle $f\mathrm{O}_2$, assuming a modern Earth-like flux of magma to the surface. The shaded areas indicate that \ce{H2} or \ce{CO2} domination was achieved within 1\,Gyr. Venus is plotted at the BSE H/C ratio, over a range of mantle $f\mathrm{O}_2$'s suggested for the ancient and modern Earth \cite{bezos2005Fe3SFeRatios, aulbach2016EvidenceReducingArchean}as a reference value. The dashed lines for \ce{H2} indicate bounds (thinner than the linewidth at higher H/C) where \ce{H2O} is dominant at 10\,Gyr, but \ce{H2} has previously been the dominant species. The lower bound acts as the 10\,Gyr \ce{H2} dominant contour.}
    \label{fig:contours}
\end{figure}

The chemical composition of volcanic atmospheres is expected to slowly change through time as the surface pressure of the planet increases, driving preferential degassing of the most insoluble species \citeA<e.g., carbon>[]{gaillard2014TheoreticalFrameworkVolcanic}. Three examples of long-term atmospheric evolution are shown in Fig.\,\ref{fig:through_time}. As previously discussed in Sects.\,\ref{section:fo2-classes} and \ref{section:hc_ratio}, we find that in most cases atmospheric chemical compositions have stabilised with respect to the classes we present here after 1-3\,Gyr of volcanic outgassing (e.g., Fig.\,\ref{fig:through_time}\,b). Although compositions continue to slowly evolve beyond this age, most notably in terms of which molecule makes up the dominant atmospheric species (e.g., Fig.\,\ref{fig:through_time}\,b shows \ce{CO2} becoming dominant after $\sim$\,6\,Gyr, see Fig.\,\ref{fig:contours} for more detail), class boundaries are not crossed.

However, for planets which sit close to a class boundary after 3\,Gyr (see Fig.\,\ref{fig:hc_ratio}\,a), the speciation of the atmosphere can change more dramatically over long timescales (Figs.\,\ref{fig:through_time}\,a \& \ref{fig:through_time}\,c). Our results show model atmospheres on the Class I - O boundary which start in Class O can transition into Class I over long timescales from volcanic degassing alone (i.e., discounting atmospheric loss and volatile cycling processes; Fig.\,\ref{fig:through_time}\,c). Similarly, atmospheres on the R-I boundary categorised as Class R after 3\,Gyr can transition into Class I over time (Fig.\,\ref{fig:through_time}\,a); although given the two criteria for a Class R atmosphere (COS$<$1\,$\times10^{-6}$ and CO$<$\ce{NH3}) this transition can be much more gradual, over several Gyr, in comparison to the transition shown in Fig.\,\ref{fig:through_time}\,c. It is important to note that despite the apparent sharp change in atmospheric composition seen particularly clearly in Fig.\,\ref{fig:through_time}\,c, there is no sudden change in the gasses being added to the atmosphere (Fig.\,\ref{fig:through_time}\,d-f). Over time, the pressure-sensitive nature of degassing means that sulfur and then water degassing is limited, so the slowly changing ratios of elements in the atmosphere seen in Figs.\,\ref{fig:through_time}\,d-f eventually trigger a change in the thermochemical equilibrium balance of the atmosphere so that more reduced species are favoured.

Figure\,\ref{fig:contours} shows the length of time a planet has to be volcanically active for, at a constant rate, in order for the atmosphere to become dominated by a species other than water. \ce{H2O} is much more soluble than either carbon-bearing species, or \mbox{\ce{H2}}. As the surface pressure increases over time, volcanic gases become increasingly dry, leading to atmospheres which gradually become either \ce{H2} or \ce{CO2} dominated, as seen in Figs.\,\ref{fig:through_time}\,b and \ref{fig:through_time}\,c, with the shaded areas of Fig.\,\ref{fig:contours} indicating a non-\ce{H2O} species becoming dominant in $<$\,1\,Gyr. In the case of highly reduced mantles ($ \lesssim $\,IW-2) the low $f\mathrm{O}_2$ means \ce{H2O} is never the dominant species after a few time-steps, as most of the H in the mantle is stored and outgassed as \ce{H2}.

The timescale water replacement occurs over is a function of the mantle $f\mathrm{O}_2$, the bulk silicate H/C ratio, and the melt flux to the surface (i.e., the intensity of volcanic activity on a planet). The timescales for a given species to achieve dominance shown in Fig.\,\ref{fig:contours} are calculated using a single melt flux to the surface (1\,$\times10^{14}$\,kg\,yr$^{-1}$, similar to that of the modern Earth); the timings shown in Fig.\,\ref{fig:contours} are inversely proportional to this melt flux. This means a planet which is twice as volcanically active as Earth, with a H/C ratio of 1 and a mantle $f\mathrm{O}_2$ of IW+3 would attain a \ce{CO2} dominated atmosphere in 1\,Gyr, rather than the 2\,Gyr shown in Fig.\,\ref{fig:contours}.

For higher H/C ratios, a carbon-rich atmosphere is never achieved even under oxidised scenarios. This is best understood by imagining a simple case where the only volatiles considered are \ce{H2O} and \ce{CO2}. Under high H/C conditions, the overabundance of H is such that even accounting for the greater fraction of C which can reach the atmosphere (due to it's lower solubility; in our simulations $>$\,90\% of the H remains in the mantle after 10\,Gyr of outgassing at IW+4, compared to 18\% of the C), the amount of water extracted from the mantle will be greater than the amount of \ce{CO2}, and a \ce{CO2} dominated atmosphere will not occur by closed system volcanic degassing alone.

\section{Discussion}
\label{section:discuss}

We have identified three atmospheric classes for hot, volcanically-derived atmospheres. These apply for planets where minimal escape or volatile cycling is occurring. These classes are robust to a wide range of water-rich to water-poor volatile inventories and show that planetary interiors, specifically the mantle $f\mathrm{O}_2$, can impose constraints of the chemistry of a planet's atmosphere even after significant temperature changes and thermochemical re-equilibration is applied.

The presence of distinct groups in atmospheres with 800\,K surface temperatures, which are not present at 1850\,K and above, suggests that planets which are cool enough to have a solid surface, rather than a magma ocean, would be more amenable to having their mantle $f\mathrm{O}_2$ characterised. All three groups have species markers at abundances above 1\,$\times10^{-4}$/100\,ppm (\ce{SO2} for Class O, \ce{CH4} combined with \ce{CO2} for Class I, and \ce{CH4} combined with a low mean molecular weight indicating substantial \ce{H2} for Class R) which would put them above the detectability limit for the James Webb Space Telescope (JWST) \cite<e.g.,>[]{batalha2018StrategiesConstrainingAtmospheres, krissansen-totton2019AtmosphericDisequilibriumExoplanet}. However, directly applying these model atmospheres to exoplanet observations is only valid under specific scenarios, i.e. where a planet has lost it's primordial/primary atmosphere and is building a secondary atmosphere back through volcanism in a low-escape environment. Even under these constraints, uncertainties remain over atmosphere-surface interactions and photochemistry \cite{jordan2021PhotochemistryVenusLikePlanets} on hot planets, which may lead to significantly different atmospheric chemistry's to those presented here.

The atmospheres of rocky planets are highly complex and subject to a range of processes which modify their chemistry (e.g., see Table\,\ref{table:mod_processes}). The atmospheric classes discussed in this work provide an important baseline from which the effects of further atmospheric processing, such as \ce{H2} escape and photochemistry, can be evaluated. We examine the effect of varying rates of hydrogen escape on our atmospheric classes in Paper III, as this process will likely have the strongest effect on the redox state of the atmosphere, potentially weakening the link between mantle and atmosphere.

\section{Conclusions}
\label{section:concs}

Atmospheric temperature has a significant impact on the chemistry of volcanic atmospheres, and should be considered during future modelling of secondary atmospheres. Hot volcanic atmospheres at 800\,K show three distinct atmospheric classes defined according to their chemical speciation and scale height. These classes are dictated by the $f\mathrm{O}_2$ of the mantle, and are robust to a wide range of bulk silicate H/C ratios. These classes may be used as a simple base for future research, exploring the effects of other processes on volcanic secondary atmospheres as produced by a range of geological conditions.

%%

%  Numbered lines in equations:
%  To add line numbers to lines in equations,
%  \begin{linenomath*}
%  \begin{equation}
%  \end{equation}
%  \end{linenomath*}

%%%%%%%%%%%%%%%%%%%%%%%%%%%%%%%%%%%%%%%%%%%%%%%%%%%%%%%%%%%%%%%%
%
%  ACKNOWLEDGMENTS
%
% The acknowledgments must list:
%
% >>>>	A statement that indicates to the reader where the data
% 	supporting the conclusions can be obtained (for example, in the
% 	references, tables, supporting information, and other databases).
%
% 	All funding sources related to this work from all authors
%
% 	Any real or perceived financial conflicts of interests for any
%	author
%
% 	Other affiliations for any author that may be perceived as
% 	having a conflict of interest with respect to the results of this
% 	paper.
%
%
% It is also the appropriate place to thank colleagues and other contributors.
% AGU does not normally allow dedications.

\section*{Open Research}
No new experimental data were generated for this work. The data used in producing 
the figures in this work can be found in \citeA{liggins2021dataset}. The EVolve model used here, including the versions of both EVo and FastChem used to produce our results, is available at \citeA{liggins_philippa_2021_5645666}, and on the GitHub repository at \url{https://github.com/pipliggins/EVolve} where new versions will also be hosted. The FastChem 2.0 model of \citeA{stock2021} is also freely available at \url{https://github.com/exoclime/FastChem}, where updates are hosted.

\acknowledgments
The authors would like to thank both anonymous reviewers for their helpful comments which have greatly improved this manuscript. PL would like to thank Daniel Kitzmann for access to, and help with, FastChem 2.0. This work was funded by the Embiricos Trust Scholarship from Jesus College, Cambridge. S.J. thanks the Science and Technology Facilities Council (STFC) for the PhD studentship (grant reference ST/V50659X/1). P.B.R thanks the Simons Foundation for support under SCOL awards 59963.

%% ------------------------------------------------------------------------ %%
%% References and Citations

%%%%%%%%%%%%%%%%%%%%%%%%%%%%%%%%%%%%%%%%%%%%%%%
%
% \bibliography{<name of your .bib file>} don't specify the file extension
%
% don't specify bibliographystyle
%%%%%%%%%%%%%%%%%%%%%%%%%%%%%%%%%%%%%%%%%%%%%%%

\bibliography{refs}

\newpage

\appendix
\section{Sensitivity testing to pressure and fraction of mantle melting}
\label{section:c_appendix}

\begin{figure}[h!]
	\begin{center}
	\includegraphics[width=\linewidth]{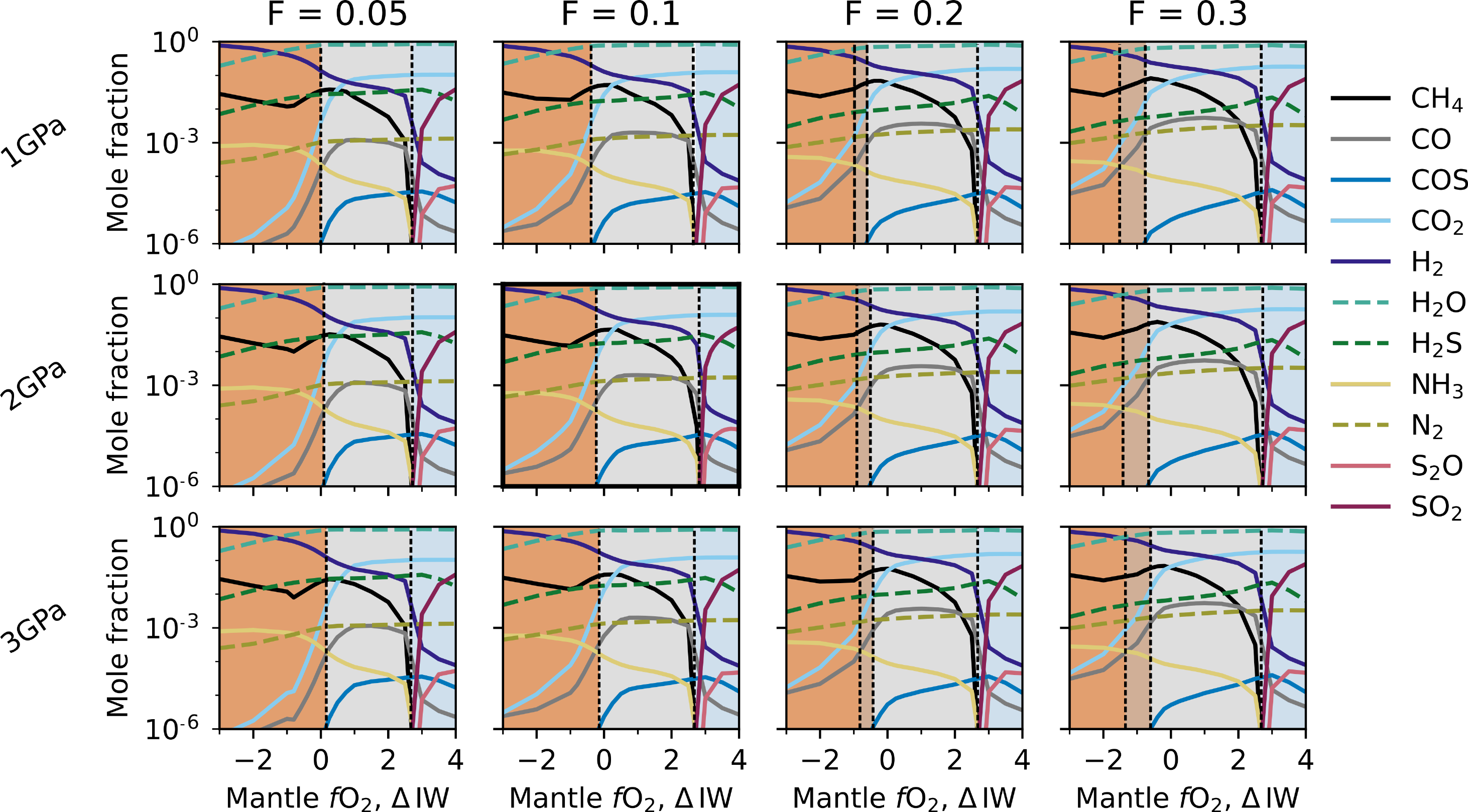}
	\caption{The sensitivity of our atmospheric classes to changes in the mantle melt fraction (F) and the pressure of mantle melting. The model discussed in the main text (F=0.1, P=2GPa) is highlighted in bold.}
	\label{fig:sensitivity}
\end{center}	
\end{figure}

Figure\,\ref{fig:sensitivity} shows the sensitivity of our class definitions to changing the pressure of mantle melting, and the local melt fraction. The model discussed in the main text is highlighted in bold. Varying the pressure of melting between 1 and 3\,GPa produces a negligible change in the mantle $f\mathrm{O}_2$ at which the class transitions occur. As the local melt fraction increases, the effect on the I-O class transition is again negligible, while it slightly increases the width of the I-R transition zone by moving the point at which \ce{NH3} becomes more abundant than CO to slightly more reduced conditions.

\section{Temperature dependence of atmospheric classes}
\label{section:t_appendix}

\begin{figure}[h!]
	\begin{center}
	\includegraphics[width=\linewidth]{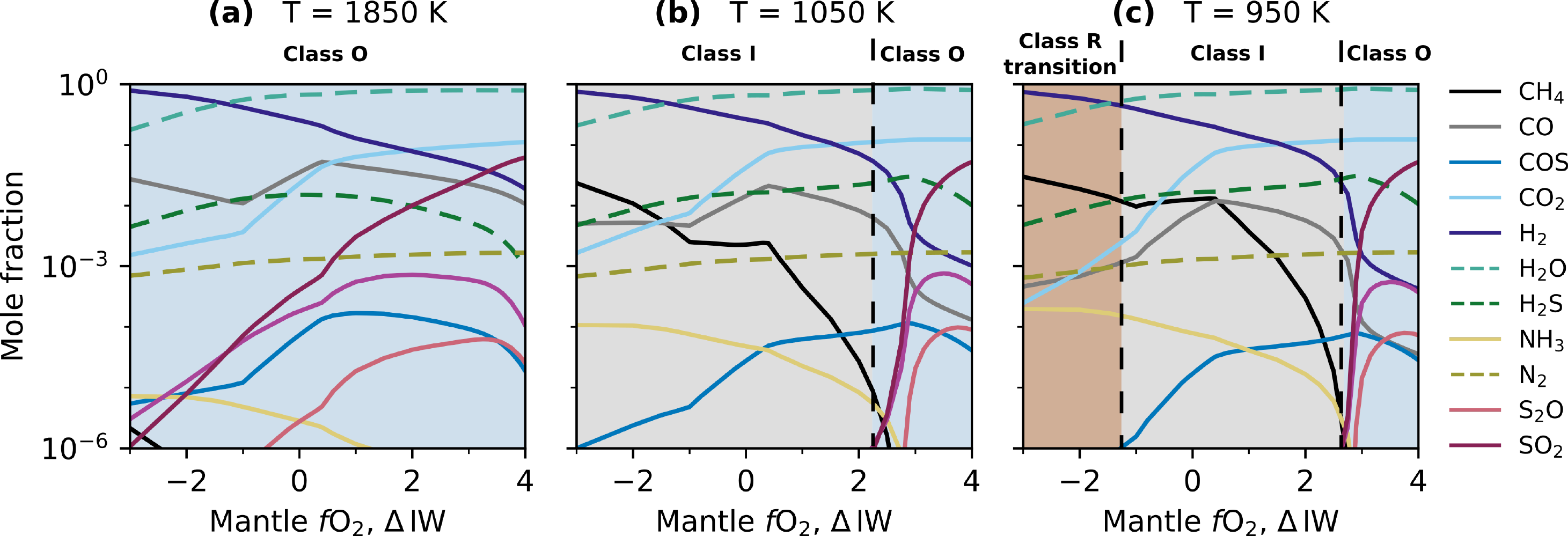}
	\caption{Atmospheric classes as the temperature is varied, used to define limits in Fig.\,\ref{fig:stellar_lines}.}
	\label{fig:appen_temps}
\end{center}	
\end{figure}

The number of atmospheric classes which are present for planets with higher surface temperatures are presented in Fig.\,\ref{fig:appen_temps}. As the surface temperature increases, Class R and then Class I atmospheres are no longer produced, regardless of mantle $f\mathrm{O}_2$. Figure\,\ref{fig:appen_temps}\,C shows that at 950\,K a transitional Class R atmosphere can be formed on planets with reduced mantles; the speciation displays almost all of the properties of a Class R atmosphere (e.g., high \ce{H2}, COS below ppm abundances), but as \ce{NH3} is not more abundant than CO it cannot be classified as a true Class R.

\end{document}